\documentclass[11pt]{article}
\usepackage{graphicx}
\usepackage{dcolumn}
\usepackage{bm}
\usepackage{amsmath}
\usepackage{amssymb}
\usepackage{tikz}
\usepackage{pgfplots}
\pgfplotsset{compat = newest}
\begin{document}

\vskip 2cm
\begin{center}

{\bf {\Large Thermodynamics of Bosonic Higher Spin Fields}}\\

\vskip 1.0cm

{\bf{\large N. Srinivas}}\\

{{\small Department of Physics, KGRCET-Hyderabad 501504, Telangana, India }} \\
{\bf{\small { { seenunamani@gmail.com}}}}
\vskip 1cm

{\bf Abstract}

\end{center}

\noindent
In this paper, I calculated the partition function of the quantum statistical system of free massless higher spin (HS) bosonic fields
with zero chemical potential value on $d$-dimensional Minkowski spacetime by using Feynman's path integral approach, which is a nontrivial result that explains the self-duality between the quantum statistical system of these fields of $s \geq 2$ over $d \geq 4$ dimensional Minkowski spacetime and the quantum statistical system of Klein-Gordon bosonic fields on 4-dimensional Minkowski spacetime at thermal equilibrium. 

However, I used the dimensional regularization approach to determine the free energy of this system, and the results indicate that the presence of ultraviolet divergences (UV) could possibly be eliminated, which also ensures that Lorentz and gauge symmetries are preserved at thermal equilibrium. However, there are still infrared (IR) divergences in this theory. And also studied the average energy and entropy of this system spanning $d \geq 4$-dimensional Minkowski spacetime. In particular, the most significant result is that the average energy spectrum of the quantum statistical system of massless free bosonic HS fields is similar to the blackbody radiation energy spectrum at different temperatures. It also explains the universe's cosmic microwave background (CMB) radiation spectrum (COBE data) at a 2.7 Kelvin temperature. The average energy spectrum of this system decreases with increasing $\beta$ value, as shown in this paper. However, the entropy of these fields also explains their self-duality with Klein-Gordon bosonic fields in 4-dimensional Minkowski spacetime.
In this paper, I studied the spontaneous symmetry breaking in the interacting bosonic higher spin fields on $d$-dimensional Minkowski spacetime, as well as the breakdown of self-duality between the interacting quantum statistical system of bosonic higher spin fields over $d\geq 4$-dimensional Minkowski spacetime and Klein-Gordon bosonic fields at thermal equilibrium points over 4-dimensional Minkowski spacetime.
\vskip 2cm

\text{\bf Keywords:}{Massless Bosonic Higher spin gauge fields, Thermal equilibrium, Entropy, 
Average energy, Quantum statistical system, Blackbody radiation, Spontaneous symmetry breaking}
 

\newpage
\section{Introduction}	

Higher spin field theory has attracted a significant amount of attention due to its unbroken phase of a string theory. Higher spin gauge theories have attracted considerable interest during the last three decades, other than being a fascinating topic by itself. Fronsdal, who has first found the equations of motion and the action principle for massless fields of arbitrary spin for $s = 1$ and also spin greater than two and also interacting higher spin fields, cannot propagate in Minkowski spacetime, which is subjected to the no-go theorem. The gauge invariance of massless free bosonic higher spin field theory has to remove nonphysical polarization and ghosts from the spectrum. However, it is highly nontrivial to study the quantum statistical behavior of massless free higher spin field theory on $d\geq 4$-dimensional Minkowski spacetime. Free massless fields have to satisfy the Klein-Gordon equation 
$\Box {\phi (x)}_{\mu_1\mu_2....\mu_s} = 0$. Here, the field would like to describe a tensor field of arbitrary rank $s$ completely symmetric in its indexes, or in other words, a field with spin $s$. Since the field is massless, it has to satisfy the $\partial^{\mu_{1}} {\phi (x)}_{\mu_1\mu_2....\mu_s} = 0$ transversality condition. However, by imposing double traceless conditions on the bosonic HS field and the traceless condition gauge parameter, which is completely symmetric, and also performing gauge symmetry transformations procedures on the action $S(\phi)$ of massless bosonic HS fields.

In this article, the symmetric massless free bosonic HS fields contain infinite sets of infinitely increasing spins over the d-dimensional Minkowski spacetime considered as a canonical ensemble of quantum statistical fields with zero chemical
potential. In particular, the most importantly calculated the partition function. for this system and got the nontrivial result that $d = 4$-dimensional Minkowski spacetime of quantum statistical systems of massless bosonic HS fields are dual to $s = 1$ and $d = 4$-dimensional. Minkowski spacetime Klein-Gordon bosonic fields. Nevertheless, when $s = 1 $and $d = 4$-dimensional Minkowski spacetime in the partition function, this system is exactly equal to the partition function of Klein-Gordon bosonic fields over $d = 4$-dimensional Minkowski spacetime

In sec. 3 However, calculated the free energy for a system of these fields on 
$d = 4$-dimensional Minkowski spacetime. This has UV divergence at one loop large momentum value, and this UV divergence could be eliminated (detached from the integral value) by using the dimensional regularization method. However, the dimensional regularization technique was introduced by Giambi and Bollini as well as independently developed by G. t Hooft and Veltman for regularizing the Feynman integral depending on the spacetime $d $ dimension.

This dimensional regularization could be used to solve the problem of divergences. Generally, this system of higher spin (HS) fields develops divergence at the loop level. The ultraviolet (UV) divergences arising at large loop momenta and also an important property of the dimensional regularization is that it respects gauge and Lorentz symmetries. Now, even after removing the UV divergences, it still has an IR divergence in this theory.

In sec. 3.1 Studied the most significant result in this paper is that the average energy distribution spectrum per unit volume of the system of massless free HS fields with chemical potential value is equivalent to zero, follows a blackbody radiation energy spectrum-like pattern across different temperatures, and also explains the Universe's cosmic microwave background radiation energy spectrum at a temperature of 2.7 Kelvin. Nevertheless, the average energy of this system decreases with increasing $\beta$ value, as shown in this paper.

In sec. 3.2 Finally, calculated the entropy of the system of these fields at thermal equilibrium, which is the most important physical quantity, such as energy.
This nontrivial result also explains the self-duality between the $s \geq 2$ of this system over $d \geq 4$-dimensional Minkowski spacetime and Klein-Gordon bosonic fields on 4-dimensional Minkowski spacetime at thermal equilibrium. 

In sec. 4 Particularly discussed the spontaneous symmetry breaking in the interacting bosonic higher spin felds and also studied
the break down of self-duality between the interacting quantum statistical system of bosonic HS fields under $d\geq 4$-dimensional Minkowski spacetimeand Klein-Gordon bosonic fields on 4-dimensional Minkowski spacetime at thermal equilibrium points.

\section{Massless Bosonic Higher spin fields}

The Fronsdal formulation of free massless bosonic higher spin fields of gauge theories (HS) [1-6] was originally formulated in terms of completely symmetric and double traceless massless bosonic HS fields $\phi(x)_{\mu_1\mu_2...\mu_s}$ (which is analogous to the metric-like formulation of gravity). In Minkowski spacetime $R^{d-1,1}$, the bosonic spin $s$ Fronsdal action is
\begin{eqnarray}
S(\phi) & = & \frac{1}{2}\int d^{d} x \Big(\partial_{\nu} {\phi (x)}_{\mu_1\mu_2....\mu_s}\partial^{\nu} {\phi (x)}_{\mu_1\mu_2....\mu_s}\\ \nonumber
& - & \frac{s(s-1)}{2}\partial_{\nu} {\phi (x)}^{\lambda}_{\lambda\mu_3....\mu_s}\partial^{\nu} {\phi (x)}_{\rho}^{\rho\mu_3....\mu_s} \\ \nonumber 
&+ & s(s-1) \partial_{\nu} {\phi (x)}^{\lambda}_{\lambda\mu_3....\mu_s}\partial_{\rho} {\phi (x)}^{\nu\rho\mu_3....\mu_s} \\ \nonumber
& - & s \,\partial_{\nu} {\phi (x)}^{\nu}_{\mu_2....\mu_s}\partial_{\rho} {\phi (x)}^{\rho\mu_2....\mu_s} \\ \nonumber
& - & \frac {s(s-1)(s-2)} {4} \partial_{\nu} {\phi (x)}^{\nu\rho}_{\rho\mu_2....\mu_s}\partial_{\lambda} {\phi (x)}^{\lambda\sigma}_{\sigma\mu_2....\mu_s}\Big)
\end{eqnarray}
where the metric like field is double traceless ($\eta^{\mu_1\mu_2}\eta^{\mu_3\mu_4}{\phi (x)}_{\mu_1\mu_2....\mu_s} = 0$)(here spin $s$ massless bosonic HS fields satisfy the traceless conditions i.e. ${\phi (x)}^{\lambda}_{\lambda\mu_3....\mu_s} = 0$ and ${\phi (x)}^{\nu\rho}_{\rho\mu_2....\mu_s} = 0$) and has the dimension of $\text{(length)}^{1-d/2}$. This action is invariant under Abelian HS gauge transformations

\begin{eqnarray}
\delta {\phi }_{\mu_1\mu_2....\mu_s}=\partial_{\mu_1} {\epsilon}_{\mu_2\mu_3....\mu_s} 
\end{eqnarray}
where gauge parameter $\epsilon$ is completely symmetric and traceless rank (s-1) tensor
\begin{eqnarray}
\eta^{\mu_1\mu_2}{\epsilon}_{\mu_1\mu_2....\mu_{s-1}} = 0. 
\end{eqnarray}

By using the above gauge symmetry transformations, the double traceless condition of the massless bosonic HS field, and the traceless condition of the gauge parameter, the Euler-Lagrange equations of motion of massless free bosonic HS fields could be obtained, which is nothing but a Klein-Gordon equation.   
\begin{eqnarray}
\Box {\phi}_{\mu_1\mu_2.....\mu_s} = 0.
\end{eqnarray}
The massless bosonic HS field would like to describe as a tensor field of arbitrary rank $s$, which is completely symmetric in its indexes.
However, if the field is massless, it has to satisfy the transversality condition. Furthermore, in order to not have propagation of states with negative norms (ghosts), the massless free bosonic HS fields should satisfy the transversality condition, i.e.,
\begin{eqnarray}
\partial^{\mu_1}\phi_{\mu_1\mu_2.....\mu_s}(x) = 0.
\end{eqnarray}
In this description, $\eta^{\mu\nu} = \text{dig}(-, +,+,.....+)$ are the components of the Minkowski spacetime metric tensor  
and $s = 1,2,3,...$ are the massless bosonic HS fields spin values.   

\section{Quantum Statistical System}
In particular, the system of massless free bosonic higher spin fields could be considered a canonical ensemble of the quantum statistical system with zero chemical potential value, which is defined over $d \geq 4$-dimensional Minkowski spacetime provided at the thermal equilibrium position. However, by using the path integral representation [11, 12], the partition function [10] of this system at inverse temperature $\beta = \frac {1}{k_B T}$ is as follows:
\begin{eqnarray}
Z(\beta) = \int {\cal D \phi} \exp\Bigl({-\beta S(\phi)}\Bigr) = \text {Tr}(\exp (-\beta S(\phi))).
\end{eqnarray}
In this context, Boltzmann's constant $k_{B} = 1$ and Planck's constant $h = 1$ are considered.
By substituting the $S(\phi)$ from (1) in the above equation (6), and this could be integrating over all massless free bosonic higher spin (HS) fields on $d$-dimensional Minkowski spacetime, then the result
as follows
\begin{eqnarray}
Z(\beta) = \int {\cal D \phi} \exp\Bigl(-{\frac{1}{2}\beta s \int d^{d} x (\phi^\nu \partial_\nu \partial_\rho \phi^\rho)}\Bigr)
\end{eqnarray}
Therefore, the result of the above Gaussian integral is as follows:
\begin{eqnarray}
Z(\beta) = (\det \,\beta s\,\Box)^{-\frac {1}{2}}.
\end{eqnarray}
This partition function result is nontrivial; it explains the physical behavior of all higher spins $s = 1, 2, 3, ....$ and when spin s = 1, it is equivalent to the partition function of the quantum statistical system of the Klein-Gordon massless free bosonic (scalar) field theory [13].

Therefore, this has been explained: there is a self-duality between the quantum statistical system of massless free bosonic HS fields of $d \geq 4$ dimensional Minkowski spacetime and Klein-Gordon bosonic fields over 4-dimensional Minkowski spacetime. This duality motivated by the concept of AdS/CFT correspondence [7, 8, 9] principle. This correspondence is the $d+1$-dimensional bulk gravitational theory (string theory) dual to the $d$-dimensional conformal field theory at the boundary value of the system.\\
Specifically, the free energy of a quantum statistical system of massless free bosonic Higher spin fields under thermal equilibrium conditions is as follows: 
\begin{eqnarray}
F = -\frac{1}{\beta}\log Z
\end{eqnarray}
here by substituting the $Z$ value from (8) in the expression (9) and obtained as follows:
\begin{eqnarray}
F(\beta) & = &\frac{1}{2\beta} \,\Bigl(\log \beta s +  \text {Tr} \log \Box \Bigr)  \nonumber \\
& = &\frac{1}{2\beta} \,\Bigl(\log \beta s +V \int \frac {d^d k} {(2 \pi)^d} \log (k^2 + \omega ^2) \Bigr) 
\end{eqnarray}
where $ \omega = \frac {2\pi}{\beta}$ is the angular frequency of massless bosonic HS fields. This free energy has an infinitely large momentum value $k$, i.e., $k \to \infty$ at short distance scales; therefore, this could be a UV divergence. However, the dimensional regularization technique [15, 16] could be used in equation (10). Now, the equation (10) becomes as follows:
\begin{eqnarray}
F(\beta) = \frac{1}{2\beta}\Bigl(\log \beta s - \frac{V \,\Gamma (-\frac {d}{2})} {{(4 \pi)}^{\frac{d}{2}}} \,\,{(\omega^2)}^{\frac{d}{2}} \Bigr)
\end{eqnarray}
And by using this method, the ultraviolet (UV) divergence could be eliminated at the one loop level, ensuring that this theory
respects standard higher spin gauge and Lorenz symmetries, though infrared (IR) divergences persist in flat spacetime configurations;  
by substituting $ \omega = \frac {2\pi}{\beta}$ in the above equation (11), then the free energy is equivalent to 
\begin{eqnarray}
F (\beta) = \frac{1}{2\beta}\bigl(\log \beta s - \frac{V \,\Gamma (-\frac {d}{2})} {{(4 \pi)}^{\frac{d}{2}}} \,\,
{(\frac {2\pi}{\beta})} ^{d} \bigr).
\end{eqnarray}
Here $V = L^d \beta$, where $L^d$ is the volume of $d$-dimensional Minkowski spacetime and the factor $\Gamma (\frac{-d}{2})$ is undefined at $d = 4, 6, 8, ...$ because the Gamma function is undefined (has poles) at zero and the negative integer values. So, to get a finite result, the Gamma function for negative integers [17] can be defined as follows:
\begin{eqnarray}
\Gamma (-r) = \frac {(-1)^r}{r!} \phi (r) - \frac {(-1)^r}{r!} \gamma \nonumber
\end{eqnarray}
where $r = 1, 2, 3, ...$,$\phi(r) = \sum_{i=1}^{r}\frac{1}{i}$ and $\gamma $ denotes Euler's constant which is defined as follows
\begin{eqnarray}
\gamma = \lim_{n\to \infty} \sum_{k=1}^{n}\Bigl(\frac{1}{k} - \ln (n)\Bigr) \nonumber
\end{eqnarray}
Since this free energy is an extensive quantity and diverges in the thermodynamic limit, $V \to \infty$. So, in this theory, infrared divergences are present even after the dimensional regularization method. Nevertheless, these infrared (IR) divergences are actually not a problem because they scale with the volume of spacetime. However, after the dimensional regularization of the free energy of these fields, it is a finite value when the volume, temperature, and spin values are finite in the dimensional regularization scheme; the logarithmic UV divergences of these fields obtain completely detached in this case. \\

\subsection{Average Energy of Bosonic Higher Spin Fields}
In principle, the average energy of the above-discussed system at the thermal equilibrium condition is expressed as
\begin{eqnarray}
E & = & -\frac{\partial }{\partial \beta} \log Z  \nonumber \\
 & = & F + \beta \frac {\partial F} {\partial \beta}. 
\end{eqnarray}
In the above equation (13), by substituting the free energy value of equation (12) and after calculation, the average energy result is as follows:
\begin{eqnarray}
E(\beta) =  \frac{1}{2\beta}+ \frac{V \,d\,\Gamma (-\frac {d}{2})} {{2 \beta\,(4 \pi)}^{\frac{d}{2}}} \,\,{(\frac {2\pi}{\beta})}^{d}.
\end{eqnarray}
This result shows that the average energy of this system is independent of spin $s$ of the fields, and it is not in agreement with the equipartition theorem [14]; hence, it deviates from the classical nature of fields; therefore, it is the key signature of the quantum statistical nature of massless free bosonic higher spin fields with zero chemical potential value. However, the average energy distribution spectrum per unit volume versus $\beta$ by using the equation (14) over four-dimensional Minkowski spacetime is as follows. 
\begin{center}
\begin{tikzpicture}
\begin{axis}[title = \small {Fig. 1: Average energy spectrum per unit volume of a 
Quantum statistical system of massless bosonic HS fields },
xlabel = $\beta $,
ylabel = $E(\beta)$]
\addplot[domain = 50: 5]{1/x + 1/x^5};
\end{axis} 
\end{tikzpicture}
\end{center}
However, this spectrum could explain how $E(\beta )$ depends on the absolute temperature of the system, and also from this spectrum, the average energy diverges when $\beta \to 0 $  $(T \to \infty)$ high-temperature limit, but the early Universe was extremely hot and dense with a temperature is estimated to have reached the maximum possible limit known as the Planck temperature which is reaching approximately $1.4 \times 10^{32}$ Kelvin.As the universe expanded rapidly and cooled down, its temperature dropped significantly across different timeframe. Nevertheless, at this temperature the average energy of this system was maximum, and when $\beta \to \infty$, i.e., the temperature approaches the zero value, then the average energy spectrum of these fields approaches the zero value. Therefore, the average energy of this system decreases with increasing $\beta$ value, as shown in figure 1.
Nevertheless, a very interesting thing is that this average energy value does not depend on the spins of these fields. In particular, the most significant result is that this spectrum is similar to the blackbody energy spectrum at different temperature values, i.e., Fig. 1 spectrum could satisfy the spectrum obtained from the Max Planck blackbody radiation energy distribution per unit volume in thermal equilibrium [18, 19] at different temperature values by the following relation:

\begin{eqnarray}
E(\lambda, T) = \frac{8\pi h c}{\lambda^5\Bigl(\exp(\frac{hc}{k_B T\lambda }) - 1\Bigr)} \nonumber
\end{eqnarray}

In this context, $\lambda$ represents the wavelength, $h$ stands for Planck's constant, $c$ denotes the speed of light, and $T$ indicates the absolute temperature associated with blackbody radiation. Specifically, the wavelength at which the blackbody radiation curve peaks serves as an indicator of temperature and illustrates the relationship between emitted radiation energy and temperature. It is possible that this field system is integral to the universe's evolution. Consequently, the early Universe would have consisted solely of the blackbody radiation from these fields. A crucial insight provided by cosmological observations is that the universe cooled due to its expansion, while still maintaining the characteristics of blackbody thermal radiation. However, figure 1 can also clarify the average energy of these fields is at a temperature close to 2.7 Kelvin, which aligns with the energy spectrum of the universe's cosmic microwave background radiation (CMB) at this temperature[20], as shown by experimental data (COBE) indicating that the CMB radiation spectrum closely resembles the energy spectrum of blackbody radiation. The thermal characteristics of blackbody radiation from these fields were crucial for comprehending the formation of the Universe's early matter[21].\\

\subsection{Entropy of Bosonic Higher Spin Fields }
In particular, from the quantum statistical field theory perspective, entropy is a physical quantity as fundamental as energy; therefore, the entropy of a quantum statistical system of massless free bosonic higher spin fields calculated over $d$-dimensional Minkowski spacetime is as follows:
\begin{eqnarray}
S(\beta) & = & -\frac{\partial F} {\partial T}  \nonumber \\ 
& = & \log Z +  \beta E =  \beta^2 \,\frac{\partial F} {\partial \beta} 
\end{eqnarray}
Here, by substituting the free energy value of Eq. (12) in the above equation (15), one could get the entropy result as follows:
\begin{eqnarray}
S(\beta) =  \frac{1}{2}\bigl(1 -  (\log \beta s  )\bigr ) + \frac{ V \,\Gamma (-\frac {d}{2})} {{2\,(4 \pi)}^{\frac{d}{2}}} \,\,{(\frac {2\pi}{\beta})}^{d}( 1+ d ). 
\end{eqnarray}
Therefore, the above result shows that the entropy of the quantum statistical system of these fields diverges as the temperature $\beta \to 0$ or $T \to \infty$ and also diverges at the thermodynamic limit $V \to \infty$; in this case, the entropy attains a maximum value. However, it gives a finite value at a finite value of temperature and a finite value of volume of the system, but it also depends on the spin ($s$) of these fields and also depends on the dimensionality of the system. However, by substituting $s=1$ and $d=4$ in equation (16), then the entropy is as follows:
\begin{eqnarray}
S(\beta) =  \frac{1}{2}\bigl(1 -  (\log \beta   )\bigr ) + \frac{ 5\,V \,\Gamma (-2)} {{2(4 \pi)}^{\frac{3}{2}}} \,\,{(\frac {2\pi}{\beta})}^{4}. 
\end{eqnarray}
Nonetheless, the expression (17) corresponds to the entropy of a quantum statistical system involving Klein-Gordon bosonic fields across 4-dimensional spacetime. Thus, this significant finding also clarifies the self-duality between equation (16) when $s \geq 2, d\geq 4$ and equation (17). Furthermore, as the values of spin, temperature, and dimensionality of these fields rise, their entropy value also increases, reaching its maximum when the system approaches thermal equilibrium.

\section{Interacting Theory}
The study of the interaction of quantum statistical fields of bosonic higher spin fields is extremely significant; nevertheless, the current discussion focuses solely on interacting quantum statistical fields of massless bosonic higher spin fields. However, this interaction theory is critical for comprehending the early Universe's phase transitions; at high temperatures, these fields remain symmetric and massless. At extremely high temperatures in the universe all fundamental forces (strong, electromagnetic, electroweak, and gravitational) are united. However, Bose-Einstein condensation is impossible at high temperatures because particle number is not conserved at high temperatures of this theory and also the spontaneously broken symmetries of gauge theories are restored and making particles massless. But for temperatures above the order of $T>1$ TeV, all the particles become massless.

As the Universe cools, many theories experience spontaneous symmetry breaking [22] and the fields settle into lower energy vacuum states, giving particles mass through mechanisms such as the Higgs field theory[23-25]. and also symmetry broken among all the 
fundamental forces. In principle they can interact among themselves to come to thermal equilibrium. Once it has reached thermal equilibrium, the interactions can be turned off gradually.

In principle, the interacting bosonic spin $s$  Fronsdal action over $d$-dimensional Minkowski spacetime is

\begin{eqnarray}
S_B(\phi) & = & \frac{1}{2}\int d^{d} x \Big(\Big(\partial_{\nu} {\phi (x)}_{\mu_1\mu_2....\mu_s}\partial^{\nu} {\phi (x)}_{\mu_1\mu_2....\mu_s}\\ \nonumber
& - & \frac{s(s-1)}{2}\partial_{\nu} {\phi (x)}^{\lambda}_{\lambda\mu_3....\mu_s}\partial^{\nu} {\phi (x)}_{\rho}^{\rho\mu_3....\mu_s} \\ \nonumber 
&+ & s(s-1) \partial_{\nu} {\phi (x)}^{\lambda}_{\lambda\mu_3....\mu_s}\partial_{\rho} {\phi (x)}^{\nu\rho\mu_3....\mu_s} \\ \nonumber
& - & s \,\partial_{\nu} {\phi (x)}^{\nu}_{\mu_2....\mu_s}\partial_{\rho} {\phi (x)}^{\rho\mu_2....\mu_s} \\ \nonumber
& - & \frac {s(s-1)(s-2)} {4} \partial_{\nu} {\phi (x)}^{\nu\rho}_{\rho\mu_2....\mu_s}\partial_{\lambda} {\phi (x)}^{\lambda\sigma}_{\sigma\mu_2....\mu_s} \Big) \\ \nonumber
& + & V(\phi)\Big)
\end{eqnarray}
where 
\begin{eqnarray}
V(\phi) &= &\frac{\mu^2}{2} \phi^2 + \frac {\lambda}{4!} \phi^4 \\ \nonumber
              & = &\frac{\mu^2}{2} \, \phi_{\mu_1\mu_2....\mu_s} \phi^{\mu_1\mu_2....\mu_s} + \frac{\lambda}{4!}  (\phi_{\mu_1\mu_2....\mu_s} \phi^{\mu_1\mu_2....\mu_s})^2 \\\nonumber 
\phi^2 & = &(\phi_{\mu_1\mu_2....\mu_s} \phi^{\mu_1\mu_2....\mu_s}) \\ \nonumber
\phi^4 & = &(\phi_{\mu_1\mu_2....\mu_s} \phi^{\mu_1\mu_2....\mu_s})^2 
\end{eqnarray}
where $\lambda \geq 0$ and $\mu^2 > 0$ for stability of the vacuum. 
\begin{eqnarray}
S_B(\phi) = S(\phi) + \frac{1}{2}\int d^{d} x \Big(\frac{\mu^2}{2} \, \phi^2 
+ \frac{\lambda}{4!}  (\phi)^4\Big) \nonumber
\end{eqnarray}
This action is invariant under Abelian Higher spin gauge transformations
\begin{eqnarray}
\delta {\phi }_{\mu_1\mu_2....\mu_s}=\partial_{\mu_1} {\epsilon}_{\mu_2\mu_3....\mu_s} \nonumber
\end{eqnarray}
where gauge parameter $\epsilon$ is completely symmetric and traceless rank (s-1) tensor
\begin{eqnarray}
\eta^{\mu_1\mu_2}{\epsilon}_{\mu_1\mu_2....\mu_{s-1}} = 0. \nonumber
\end{eqnarray}
However, in order to not have propagation of states with negative norms (ghosts), the massless bosonic HS fields should satisfy the transversality condition, i.e.,
\begin{eqnarray}
\partial^{\mu_1}\phi_{\mu_1\mu_2.....\mu_s}(x) = 0. \nonumber
\end{eqnarray}
Therefore the gauge symmetry is there in this theory. \\
In particular, the local Lorentz symmetry transformation as follows
\begin{eqnarray}
{\phi }^{\mu_1\mu_2....\mu_s}(x) \longrightarrow \Lambda_{b}^{\mu_1\mu_2....\mu_s}(x)  \phi^b(x).   
\end{eqnarray}
Therefore, under the local Lorentz symmetry transformation of equation (18), it is not invariant. However, in order to restore the local Lorentz symmetry invariance by introducing the covariant derivatives in the action (18). \\
In principle, the interacting bosonic spin $s$  Fronsdal action in covariant derivative form on $d$-dimensional
Minkowski spacetime is 

\begin{eqnarray}
S_B(\phi) & = & \frac{1}{2}\int d^{d} x \Big(\Big(D_{\nu} {\phi (x)}_{\mu_1\mu_2....\mu_s}D^{\nu} {\phi (x)}_{\mu_1\mu_2....\mu_s}\\ \nonumber
& - & \frac{s(s-1)}{2}D_{\nu} {\phi (x)}^{\lambda}_{\lambda\mu_3....\mu_s}D^{\nu} {\phi (x)}_{\rho}^{\rho\mu_3....\mu_s} \\ \nonumber 
&+ & s(s-1) D_{\nu} {\phi (x)}^{\lambda}_{\lambda\mu_3....\mu_s}D_{\rho} {\phi (x)}^{\nu\rho\mu_3....\mu_s} \\ \nonumber
& - & s \,D_{\nu} {\phi (x)}^{\nu}_{\mu_2....\mu_s}D_{\rho} {\phi (x)}^{\rho\mu_2....\mu_s} \\ \nonumber
& - & \frac {s(s-1)(s-2)} {4} D_{\nu} {\phi (x)}^{\nu\rho}_{\rho\mu_2....\mu_s}D_{\lambda} {\phi (x)}^{\lambda\sigma}_{\sigma\mu_2....\mu_s} \Big) \\ \nonumber
& + & V(\phi)\Big)
\end{eqnarray}
where covariant derivative defined as follows
\begin{eqnarray}
D_\mu({\phi (x)}^{\mu_1\mu_2....\mu_s}) = \partial_{\mu} {\phi (x)}^{\mu_1\mu_2....\mu_s}
+ \omega_{\mu \,c_1}^{\mu_1} {\phi (x)}^{c_1\mu_2....\mu_s} 
+ .....+ \omega^{\mu_n}_{\mu \,c_n} {\phi (x)}^{\mu_1\mu_2....c_n} \nonumber
\end{eqnarray}
\begin{eqnarray}
\phi^a \longrightarrow \Lambda_{b}^a \phi^b \nonumber
\end{eqnarray}
\begin{eqnarray}
D_\mu{\phi (x)}^{\mu_1\mu_2....\mu_s}\longrightarrow \Lambda_{b}^{\mu_1\mu_2....\mu_s}(x) D_\mu  \phi^b(x) \nonumber
\end{eqnarray}
\begin{eqnarray}
D_\mu \phi^b(x) = \partial_\mu \phi^b(x) + \omega_\mu^{b} a \,\phi^a \nonumber
\end{eqnarray}
\begin{eqnarray}
\omega_\mu^{b} a = e^b_\nu\,\partial_\mu E^\nu_{a} + \Gamma^\mu_\nu \rho \, e^b_{\nu} \, E^\rho_{a}  \nonumber
\end{eqnarray}
where $e^b_\nu$ is vielbein field
\begin{eqnarray}
\Gamma^{\sigma}_{\mu\,\nu} = \mathrm{e}^{\rho}_{j} \, \mathrm{e}^{j\sigma}\Bigl(e_{\rho\,I}\,\partial_{\small(\mu }\mathrm{e}_{\nu \small)}^I
+ \mathrm{e}_{\nu\,I}\,\partial_{\small(\mu } \mathrm{e}_{\rho \small)}^I + \mathrm{e}_{\nu\,I}\,\partial_{\small(\nu } \mathrm{e}_{\rho \small)}^I \Bigr) \nonumber
\end{eqnarray}
\begin{eqnarray}
e^b_\nu \longrightarrow \Lambda^b_{d} \,e^d_{\nu} \nonumber \\
E^\nu_{a} \longrightarrow E^{\nu}_c\,(\Lambda^{-1 })^{c}_{a} \nonumber
\end{eqnarray}
Minkowski solution
\begin{eqnarray}
e^I_{\mu} = \delta^I_\mu , \,\, \omega^I_{\mu J}(x) = 0  \nonumber
\end{eqnarray}
\begin{eqnarray}
\eta^{I J} e^\mu_{I}(x) e^\nu_{J}(x) =  \eta^{I J} \delta^\mu_{I}(x) \delta^\nu_{J}(x) = \eta^{\mu\nu} \nonumber
\end{eqnarray}
Here, the $ \eta^{\mu\nu}$ Minkowski metric of special theory of relativity is nothing but a particular value of the gravitational field.
However, When $\mu^2 < 0$, the interacting bosonic higher-spin fields undergo spontaneous symmetry breakdown due to quantum fluctuations in their vacuum state of dynamic fields within the system, as described in equation (21). In principle, after spontaneous symmetry breaking, vector gauge bosons gain mass via the Higgs mechanism; similarly, after spontaneous symmetry breaking in the bosonic higher-spin fields, a tower of bosonic higher-spin mass particles could be produced via the Higgs mechanism, i.e., this is the broken phase of the bosonic higher-spin fields. However, after spontaneous symmetry breakdown in the system causes a quantum phase transition from the unbroken phase to the broken phase of the interacting bosonic HS fields.

The partition function of the quantum statistical system of interacting bosonic higher spin fields of gauge theories over $d \geq 4$-dimensional Minkowski spacetime at thermal equilibrium position by Feynman's path integral method is as follows.
\begin{eqnarray}
Z_B(\beta) = \int {\cal D \phi} \exp\Bigl({-\beta (S_B(\phi)  }\Bigr) = \text {Tr}(\exp (-\beta S_B(\phi)))
\end{eqnarray}
by substituting $S_B$ from the equation (18) in the above equation, then

\begin{eqnarray}
Z_B(\beta) = \int {\cal D \phi} \exp\Bigl({-\beta \Bigl(S(\phi) + \frac{1}{2}\int d^{d} x \Big(\frac{\mu^2}{2} \, \phi^2 
+ \frac{\lambda}{4!}  (\phi)^4\Bigr) }\Bigr) \Bigr)
\end{eqnarray}

\begin{eqnarray}
Z_B(\beta) = \int {\cal D \phi} \exp\Bigl({-\beta \Bigl(S(\phi)}\Bigr)
\cdot \int {\cal D \phi} \exp\Bigl( \frac{1}{2}\int d^{d} x \Big(\frac{-\beta\mu^2}{2} \, \phi^2 \Bigr) 
\cdot \int {\cal D \phi} \exp\Bigl( \sum^{\infty}_{N = 0} \frac{1}{N!} (-\beta\,S_I)^N \Bigr)
\end{eqnarray}
where
\begin{eqnarray}
 S_I = \int d^{d} x \frac {\lambda}{4!} \phi^4 \nonumber
\end{eqnarray}
\begin{eqnarray}
Z_B(\beta) = \Bigl(\det \,\beta s\,\Box\Bigr)^{-\frac {1}{2}}\cdot \Bigl(\frac{4\pi}{\beta \mu^2}\Bigr)^{\frac{1}{2}}
\cdot \int {\cal D \phi} \exp\Bigl( \sum^{\infty}_{N = 0} \frac{1}{N!} (-\beta\,S_I)^N \Bigr)
\end{eqnarray}
where 
\begin{eqnarray}
\int {\cal D \phi} \exp\Bigl( \sum^{\infty}_{N = 0} \frac{1}{N!} (-\beta\,S_I)^N \Bigr). \nonumber
\end{eqnarray}
This is equal to the integral over the exponential sum of the connected Feynman bubble diagrams of interacting bosonic HS fields, but it provides a finite value at a particular temperature of the system.
However, the result (25) shows that the breaks down the self-duality between the interacting quantum statistical system of bosonic higher spin fields over $d\geq 4$-dimensional Minkowski spacetime and Klein-Gordon bosonic fields over 4-dimensional Minkowski spacetime at thermal equilibrium points, this is because of the interaction of quantum statistical system of the bosonic higher spin fields.

\section{Conclusions}
\noindent
The Big Bang theory suggests that the universe began as a dense, extremely hot single point around 13.8 billion years ago. In its early phases, the universe was totally composed of quantum statistical fields. As a result, studying these fields is essential for understanding the universe's physical nature. 

In this article, I applied the path integral method to calculate the partition function for the quantum statistical fields of free massless bosonic higher spins with zero chemical potential value over the $d\geq 4$-dimensional Minkowski spacetime at thermal equilibrium position. 

This partition function is a nontrivial result explains the self-duality between the $s \geq 2$ system of these fields over $d \geq 4$-dimensional Minkowski spacetime and Klein-Gordon bosonic fields over 4-dimensional Minkowski spacetime at thermal equilibrium position. It also calculated the free energy of this system. I utilized the dimensional regularization technique to eliminate the UV divergences that arose in this calculation at the one loop level, while preserving Lorentz and HS gauge symmetries. However, it discussed both the average energy and the entropy of this system. In particular, the dimensional regularization technique is particularly beneficial in removing the UV divergences that emerge in these computations and obtaining nontrivial answers; nevertheless, IR divergences are still present in this theory, and they are not an issue because they scale with the volume of the spacetime. The key finding in this paper, however, is that the quantum statistical system of the average energy spectrum of these fields is comparable to the blackbody energy spectrum at various temperatures. It also explains the cosmic microwave background radiation in the universe at a temperature of 2.7 kelvin, and the average energy value decreases as the $\beta$ value increases, as this paper illustrates. Additionally, the quantum statistical entropy of these fields reveals a self-duality between the Klein-Gordon bosonic fields over 4-dimensional Minkowski spacetime at thermal equilibrium positions and the $s \geq 2$ of these fields over $d \geq 4$-dimensional Minkowski spacetime. In this paper,the quantum phase transition from unbroken to broken phase of bosonic HS field can be understood through the spontaneous symmetry breaking in the interacting bosonic higher spin fields. The breakdown of self-duality between the interacting quantum statistical system of bosonic HS fields over $d\geq 4$-dimensional Minkowski spacetime and Klein-Gordon bosonic fields on four dimensional Minkowski spacetime at thermal equilibrium points has also been considered. 

\begin{center}
\bf {Appendix A}
\end{center}
The action for Klein-Gordon massless free bosonic fields  on 4-dimensional Minkowski spacetime is
\begin{eqnarray} 
S_o(\phi) = \int d^4 x \Bigl(\frac{1}{2}(\partial_\mu \phi)^2 \Bigr)\nonumber
\end{eqnarray}
\begin{center}
The partition function of this system is
\end{center}
\begin{eqnarray} 
Z_0 = \int {\cal D \phi} \exp\Bigl(-{\frac{1}{2}\beta } \int d^{4} x (\partial_\mu \phi)^2)\Bigr ). \nonumber 
\end{eqnarray}
\begin{center}
Therefore the above Gaussian integral value is equal to
\end{center}
\begin{eqnarray} 
Z_0 (\beta) = (\det \beta\,\Box)^{-\frac {1}{2}}. \nonumber
\end{eqnarray}
Here
$\partial_\mu = \frac {\partial}{\partial x^\mu}$ 
and $\partial^2 = \eta^{\mu\nu}\partial_\mu\partial_\nu = \partial_\mu\partial^\mu = \Box$ (d' Alembertian operator)
\begin{center}
\bf {Appendix B}
\end{center}

The following dimensional regularization method applied in the above calculations
\begin{eqnarray}
\int \frac {d^d p} {(2 \pi)^d} \frac {{(p^2)}^n} {(p^2 + \omega ^2)^{\alpha }} = 
\frac{(\omega^2)^{\frac{d}{2} +n -\alpha}\, \Gamma (\alpha - n -\frac{d}{2} ) \,\Gamma (\frac{d}{2} + n)}{(4\pi)^\frac{d}{2}\,\Gamma(\frac{d}{2})\, \Gamma(\alpha)} \nonumber
\end{eqnarray}
and also used the below Gamma function relation 
\begin{eqnarray}
\Gamma(a+1) = a \,\Gamma(a) \nonumber
\end{eqnarray}

\vskip 0.4cm
\noindent
{\bf Declaration}: The author declare that no competing interests \\
{\bf Data availability statement}: My manuscript has no associated data

\end{document}